\documentclass[twocolumn, showpacs, aps, pre, superscriptaddress, amsmath, amssymb, nofootinbib]{revtex4-1}

\usepackage[T1]{fontenc}
\usepackage{lmodern}
\usepackage{calrsfs}
\usepackage{upgreek}
\usepackage[mathscr]{eucal}
\usepackage{amsmath}
\usepackage{graphicx}
\usepackage{bm}
\usepackage{color}
\usepackage[dvipsnames]{xcolor}
\usepackage{hyperref}
\hypersetup{colorlinks=true,citecolor=magenta}

\usepackage{gensymb }
\usepackage{float}

\graphicspath{ {Figures/} }

\begin{document}

\title{Multiparticle Collision Dynamics for Tensorial Nematodynamics}

\author{Shubhadeep Mandal}
\affiliation{Max-Planck-Institute for Dynamics and Self-Organization, Am Fassberg 17, 37077 G\"ottingen, Germany}
\author{Marco G. Mazza}
\affiliation{Max-Planck-Institute for Dynamics and Self-Organization, Am Fassberg 17, 37077 G\"ottingen, Germany}
\affiliation{Interdisciplinary Centre for Mathematical Modelling and Department of Mathematical Sciences, Loughborough University, Loughborough, Leicestershire LE11 3TU, United Kingdom}

\date{\today}

\begin{abstract}
Liquid crystals establish a nearly unique combination of thermodynamic, hydrodynamic, and topological behavior. This poses a challenge to their theoretical understanding and modeling. The arena where these effects come together is the mesoscopic (micron) scale. It is then important to develop models aimed at capturing this variety of dynamics. 
We have generalized the particle-based multiparticle collision dynamics (MPCD) method to model the dynamics of nematic liquid crystals. Following the Qian--Sheng theory [Phys. Rev. E \textbf{58}, 7475 (1998)] of nematics, the spatial and temporal variations of the nematic director field and order parameter are described by a tensor order parameter. The key idea is to assign tensorial degrees of freedom to each MPCD particle, whose mesoscopic average is the tensor order parameter. This new nematic-MPCD  method includes backflow effect, velocity-orientation coupling and thermal fluctuations.  We validate the applicability of this method by testing: (\emph{i}) the nematic-isotropic phase transition, (\emph{ii}) the flow alignment of the director in shear and Poiseuille flows, and (\emph{iii}) the annihilation dynamics of a pair of line defects. We find excellent agreement with existing literature. We also investigate the flow field around a force dipole in a nematic liquid crystal, which represents the leading-order flow field around a force-free microswimmer. The anisotropy of the medium not only affects the magnitude of velocity field around the force dipole, but can also induce hydrodynamic torques depending on the orientation of dipole axis relative to director field. A force dipole experiences a hydrodynamic torque when the dipole axis is tilted with respect to the far-field director. The direction of hydrodynamic toque is such that the pusher- (or puller-) type force dipole tends to orient along (or perpendicular to) the director field. Our nematic-MPCD method can have far-reaching implications not only in modeling of nematic flows, but also to study the motion of colloids and microswimmers immersed in an anisotropic medium.
\end{abstract}

\pacs{}

\maketitle

\section{Introduction}

The dynamics of liquid crystals in presence of spatially and temporally varying active or passive forces are an important topic in the field of nonequilibrium soft matter. Nematic liquid crystals posses long range molecular orientational order due to their rod-like molecules \cite{andrienko2018introduction}. This orientational order is described by two key quantities: the director and the scalar order parameter. The director represents the common axis along which the molecules tend to align, while the scalar order parameter represents the degree of molecular orientation along the director \cite{stephen1974physics, sengupta2014liquid}. The dynamics of nematic liquid crystals are challenging as compared to simple isotropic fluids due to two main aspects. First, the relaxations of the order parameter, director, and momentum in nematics take place often at different length and time scales \cite{care2005computer}. Second, the director's reorientation and fluid flow are coupled. Flow-induced deformation of fluid elements leads to reorientation of the nematic molecules and thereby affects the orientational order. Furthermore, nonuniform director field can also induce macroscopic fluid flow \cite{gennes1995physics}. These inherent complexities demand a robust mesoscopic model to obtain a more comprehensive understanding of nematodynamics.     

There exists several continuum theories of nematodynamics which describes the orientational order of nematic fluids either in terms of director field (assuming the scalar order parameter as a frozen degree of freedom) or tensor order parameter \cite{care2005computer}. 
Ericksen, Leslie and Parodi (ELP) developed a nematodynamic theory in terms of director field \cite{stewart2004static}. Though this theory successfully explains several experiments, its applicability  is limited as the scalar order parameter is assumed to be constant, which prevents it from describing systems with topological defects. 
Later, Beris and Edwards \cite{beris1994thermodynamics} and Qian and Sheng \cite{qian1998generalized} developed tensorial theories which take into account the spatial and temporal variation of both the director and the scalar order parameter. 
Both of these theories describe the nematic orientational order in terms of a tensor quantity which effectively combines the director and scalar order parameter. Researchers have solved these nematodynamic equations using the lattice Boltzmann method \cite{care2000lattice, denniston2001lattice, care2003generalized} or standard finite-difference and finite-element methods \cite{james2006finite, james2008computer} to study the nematodynamics in different forcing conditions ($e.g.$ shear flow, Poiseuille flow, defect dynamics etc. \cite{denniston2001phase, denniston2004lattice, toth2002modeling, toth2003hydrodynamics, ravnik2009landau, lintuvuori2010colloids, zhang2016lattice, giomi2017cross}). 
One limitation of these methods is the absence of thermal fluctuations. It is important to mention that there are several physical situations in which both the thermal fluctuations and hydrodynamics can play important role. For example, the dynamics of nematic colloids and nematic droplet swimmers \cite{smalyukh2018liquid, kruger2016curling}. 

The multiparticle collision dynamics (MPCD) \cite{malevanets1999mesoscopic} has been extensively used as a mesoscale simulation method with the  benefit of incorporating thermal fluctuations with long range hydrodynamics \cite{gompper2009multi, kapral2008multiparticle}. 
Initial attempts at developing a fluctuating nematodynamic model have been made by Lee and Mazza \cite{lee2015stochastic} and Shendruk and Yeomans \cite{shendruk2015multi}. Both approaches assigned an orientation vector to each MPCD particle. 
Lee and Mazza used the Lebwohl--Lasher potential to describe the interaction among particles' orientations. The velocity-orientation coupling and the backflow effect were incorporated by using macroscopic spatial derivatives of velocity gradient and elastic stress following the simplified ELP theory. On the other hand, Shendruk and Yeomans have used the Maier--Saupe potential and the Jeffery's equation to update the particle orientation, while the backflow effect was included in angular momentum balance. Though these methods have successfully reproduced several features of nematic liquid crystals ($e.g.$ nematic-isotropic phase transition, shear alignment, and defect dynamics), there is no MPCD method which describes the orientational order in terms of the tensor order parameter. 

In this work, a new MPCD scheme is formulated by implementing the equations of the Qian--Sheng theory. The Andersen-thermostatted MPCD method for isotropic fluid is extended to incorporate the orientational order of nematic liquid crystals. 
Towards this, extra degrees of freedom in terms of a tensor order parameter is assigned to each MPCD particle. The temporal evolution of this tensor order parameter is governed by the molecular field and velocity-orientation coupling terms which are incorporated by using mesoscopic derivatives. 
The orientation-velocity coupling which takes into account of the anisotropic viscous stress and elastic stress are incorporated by adding a forcing term in the streaming step of MPCD. Finally, we validate our model in different equilibrium and nonequilibrium conditions. Our results agree well with the existing literature. 

We also study the flow field and director orientation around pusher-type and puller-type force dipoles in nematic fluid. Very recently, Kos and Ravnik \cite{kos2018elementary} studied elementary flow fields ($e.g.$ Stokeslet, stresslet, rotlet, etc.) in nematic liquid crystals. Later Daddi \emph{et al.} \cite{daddi2018dynamics} studied the motion of a simple model microswimmer in nematic liquid crystals. In both of these studies, the director field and nematic order parameter were assumed spatially uniform. To go beyond the uniform-director-field approximation, here we study the velocity-orientation coupling and the effect of deformed director field on the flow field around a force dipole.

\section{Model}
\subsection{Equations of nematodynamics}
We start by briefly reviewing the nematodynamic equations of the Qian--Sheng theory \cite{qian1998generalized}. 
Qian and Sheng developed a continuum theory for nematic liquid crystals with variable order parameter. The orientational order of the nematic fluid is described in terms of a tensor order parameter $\bm{Q}$ which is traceless and symmetric. 
The tensor order parameter effectively combines the director field $\bm{n}$ and the scalar order parameter $S$. For uniaxial nematics, this can be stated as $Q_{\alpha \beta} = S (3 n_{\alpha}n_{\beta} - \delta_{\alpha \beta})/2$, where $\alpha,\beta=x,y,z$, the Cartesian coordinates. Thus, $S$ represents the largest eigenvalue of $\bm{Q}$, while the corresponding eigenvector is $\bm{n}$. 
Qian and Sheng described the nematodynamics in terms of the evolution of  $\bm{Q}$ and fluid velocity $\bm{V}$. In the limit of negligible moment of inertia density, the evolution of $\bm{Q}$ is given as \cite{qian1998generalized}      
\begin{align}\label{Q_eq}
D_{t}Q_{\alpha\beta} = & \frac{1}{\mu_{1}} H_{\alpha\beta} - \frac{\mu_{2}}{2 \mu_{1}} A_{\alpha\beta} + (Q_{\alpha\mu}W_{\mu\beta} - W_{\alpha\mu}Q_{\mu\beta}) \nonumber \\
& - \frac{1}{\mu_{1}}(\lambda\delta_{\alpha\beta} + \lambda_{\mu} \epsilon_{\mu\alpha\beta}),
\end{align}    
where $\mu_{1}$ and $\mu_{2}$ are viscosity coefficients, $D_{t} \equiv \partial_{t} + V_{\mu}\partial_{\mu}$ is the material time derivative, $A_{\alpha\beta} = \frac{1}{2}(\partial_{\alpha}V_{\beta} + \partial_{\beta}V_{\alpha})$ and $W_{\alpha\beta} = \frac{1}{2}(\partial_{\alpha}V_{\beta} - \partial_{\beta}V_{\alpha})$ are the symmetric and anti-symmetric parts of the velocity gradient tensor, respectively, $H_{\alpha\beta}$ is the molecular field, and $\lambda$ and $\lambda_{\mu}$ are Lagrange multipliers which impose traceless and symmetry conditions on $Q_{\alpha\beta}$, respectively. 
Note that the first term of the right-hand side of equation \ref{Q_eq} represents the molecular field $H_{\alpha\beta}$ which is the key to the nematic-isotropic phase transition at equilibrium, while the second and third terms represent the velocity-orientation coupling. 
The Landau--de Gennes theory gives the molecular field (assuming the one-elastic-constant approximation) as $H_{\alpha\beta} = L \partial_{\mu}^{2} Q_{\alpha\beta} - \alpha_{F} Q_{\alpha\beta} + 3\beta_{F} Q_{\alpha\mu}Q_{\beta\mu} - 4\gamma_{F} Q_{\alpha\beta}Q_{\mu\nu}Q_{\mu\nu}$, where $L$ is elastic constant and $\alpha_{F}$, $\beta_{F}$ and $\gamma_{F}$ are phenomenological material constants, and where the Einstein notation of summation over repeated indices is assumed. 

The velocity field satisfies the continuity equation, and the evolution of fluid velocity is given as \cite{qian1998generalized}   
\begin{equation}\label{V_eq}
\rho D_{t}V_{\beta}  = \partial_{\alpha} (\sigma_{\alpha\beta}^{v,\text{iso}} + \sigma_{\alpha\beta}^{v,\text{aniso}} + \sigma_{\alpha\beta}^{e}),
\end{equation}   
where $\rho$ is density of nematic fluid, $\sigma_{\alpha\beta}^{v,\text{iso}}$ is isotropic contribution to viscous stress, $\sigma_{\alpha\beta}^{v,\text{aniso}}$ is anisotropic contribution to viscous stress, and $ \sigma_{\alpha\beta}^{e}$ is elastic/distortion stress. These stresses can be expressed as \cite{qian1998generalized}
\begin{equation}\label{Vis_isoStress_eq}
\sigma_{\alpha\beta}^{v,\text{iso}} = - P \delta_{\alpha\beta} + \beta_{4}A_{\alpha\beta},
\end{equation}    
\begin{equation}\label{Vis_anisoStress_eq}
\begin{split}
\sigma_{\alpha\beta}^{v,\text{aniso}} = & \beta_{1}Q_{\alpha\beta}Q_{\mu\nu}A_{\mu\nu} + \beta_{5}Q_{\alpha\mu}A_{\mu\beta} + \beta_{6}Q_{\beta\mu}A_{\mu\alpha} \\
& + \frac{1}{2} \mu_{2}N_{\alpha\beta} - \mu_{1}Q_{\alpha\mu}N_{\mu\beta} + \mu_{1}Q_{\beta\mu}N_{\mu\alpha},
\end{split}
\end{equation}        
\begin{equation}\label{El_Stress_Eq}
\sigma_{\alpha\beta}^{e} = - L \partial_{\alpha}Q_{\mu\nu}\partial_{\beta}Q_{\mu\nu}, 
\end{equation}  
where $\beta_{1}$, $\beta_{4}$, $\beta_{5}$ and $\beta_{6}$ are viscosity coefficients, $P$ the pressure, and $N_{\alpha\beta} = D_{t}Q_{\alpha\beta} + W_{\alpha\mu}Q_{\mu\beta} - Q_{\alpha\mu}W_{\mu\beta}$ is the corotational derivative. Note that the anisotropic viscous stress and elastic stress represent the orientation-velocity coupling, also called the backflow effect.

\subsection{Simple MPCD for isotropic fluids}
Before describing the nematic-MPCD model, we first look into the key steps of the simple MPCD method which produces both long range hydrodynamic modes and thermal fluctuations \cite{malevanets1999mesoscopic,gompper2009multi}. The isotropic fluid is represented by $N$ point particles (labelled with $i$) having mass $m_{0}$, position $\bm{r}_{i}$, and velocity $\bm{v}_{i}$. Note that these particles do not represent the actual fluid molecules, rather they can be thought of as a representation of a parcel of fluid. The dynamics of MPCD particles consists of alternating streaming and collision steps. These steps are constructed such that important macroscopic quantities of interest ($e.g.$ mass, momentum and energy) are conserved. In the absence of external force, the streaming is simply the ballistic motion of particles
\begin{equation}\label{Streaming}
\bm{r}_{i}(t+\Delta t) = \bm{r}_{i}(t) + \bm{v}_{i}(t)\Delta t,
\end{equation} 
where $\Delta t$ is the time between two consecutive collision steps. The collision is a stochastic process which model the interaction among MPCD particles via momentum exchange. Here we focus on the Andersen thermostat version of MPCD which also conserves angular momentum (MPC-AT+a) \cite{gompper2009multi, gotze2007relevance, noguchi2007h}. To perform the collision step, the system is first partitioned into cubic cells (labelled $c$) of side length $a_{0}$; particles interact only with particles   in the same cell.  
In the collision step, the velocity of each particle is  randomly reassigned from a Maxwell--Boltzmann distribution so that the cell's center-of-mass velocity and angular momentum are conserved
\begin{equation}\label{Collision}
\begin{split}
{\bm{v}}_{i} (t + \Delta t) = & \frac{1}{N_{c}} \sum_{j \in \text{cell}} \bm{v}_{j} (t) + {\bm{v}}_{i}^{\rm{ran}} - \frac{1}{N_{c}}\sum_{j \in \text{cell}}{\bm{v}}_{j}^{\rm{ran}} \\
& + {\bm{\Pi}_{c}^{-1} \sum_{j \in \text{cell}}[\bm{r}_{j,c} \times (\bm{v}_{j} - \bm{v}^{\rm{ran}}_{j})] \times \bm{r}_{i,c}},
\end{split}
\end{equation}     
where $N_{c}$ is the number of particles within the cell, $\bm{v}_{i}^{\rm{ran}}$ is a random velocity sampled from the Maxwell--Boltzmann distribution with standard deviation $\sqrt{k_{B}T_{0}/m_{0}}$ and zero mean,  $\bm{\Pi}_{c}$ is moment of inertia tensor of the cell, $\bm{r}_{j,c} = \bm{r}_{j} - \bm{r}_{c}$ is relative position of particle $j$ in the cell, and $\bm{r}_{c}$ is center of mass of the cell. 
In this way the mass, linear momentum and angular momentum are conserved locally ($i.e.$ in each collision cell), and these simple rules in long length and time limit reproduce the Navier--Stokes behavior for Newtonian fluids with thermal fluctuations. Note that a random shift of the collision-cell grid is necessary to impose the Galilean invariance as it is destroyed by the partition of the system into cells \cite{ihle2001stochastic}.

\subsection{ Nematic-MPCD for nematic fluids}
Now, the challenge of developing a nematic-MPCD model, which can reproduce the nematodynamic equations of the Qian--Sheng theory, lies in representing the evolution of the tensor order parameter and the backflow effect (represented by the aniosotropic and elastic stresses) within the present particle-based framework. Towards this end, we first augment the degrees of freedom of each MPCD particle. In addition to position and velocity, each particle is also assigned a tensor order parameter $\bm{q}$ to represent the nematic nature of the fluid. Thus, by definition $\bm{q}_{i}$ represents the tensor order parameter of a parcel of fluid represented by the $i$th MPCD particle. The key question is how to update $\bm{q}_{i}$ so that the evolution equation of $\bm{Q}$ [Eq.~\eqref{Q_eq}] is reproduced.
First, we look into the relationship between particle-based quantities and the macroscopic (or collision-cell level) quantities. The macroscopic velocity is related to the particle velocities as $\bm{V}_{c}(t) = \frac{1}{N_{c}(t)} \sum_{j \in \text{cell}} \bm{v}_{j} (t)$, while the macroscopic tensor order parameter is related to the particle-based tensor order parameter as $ \bm{Q}_{c}(t) = \frac{1}{N_{c}(t)} \sum_{j \in \text{cell}} \bm{q}_{j} (t)$. Now, we propose the following scheme to update $\bm{q}_{i}$
\begin{equation}\label{Q_update_Eq}
\bm{q}_{i}(t + \Delta t) = \bm{q}_{i}(t) +  \bm{g}_{i}(t) \Delta t,
\end{equation}  
where $\bm{g}_{i}$ is a collision-cell level tensor quantity with components 
\begin{align}
 g_{\alpha\beta} = &\frac{1}{\mu_{1}} H_{\alpha\beta} - \frac{\mu_{2}}{2 \mu_{1}} A_{\alpha\beta} + (Q_{\alpha\mu}W_{\mu\beta} - W_{\alpha\mu}Q_{\mu\beta}) \nonumber \\
 &- \frac{1}{\mu_{1}}(\lambda\delta_{\alpha\beta} + \lambda_{\mu} \epsilon_{\mu\alpha\beta}).     
\end{align}
The calculation of $g_{\alpha\beta}$ is carried out by using central difference discretization scheme across cells. 

To model the backflow effect, the key question is how to incorporate the effects of anisotropic viscous stress and elastic stress. Note that the momentum equation in nematodynamics [Eq.~\eqref{V_eq}] contains both viscous (isotropic and anisotropic) and elastic stresses. As the isotropic contribution of the viscous stress is intrinsically reproduced by the simple MPCD algorithm, the backflow effect can be incorporated by modifying the streaming step in the following way
\begin{equation}\label{Mod_Streaming_r}
\bm{r}_{i}(t+\Delta t) = \bm{r}_{i}(t) + \bm{v}_{i}(t) \Delta t +  \bm{f}_{i}(t) \frac{\Delta t^{2}}{2m_{0}},
\end{equation}
\begin{equation}\label{Mod_Streaming_v}
\bm{v}_{i}(t+\Delta t) = \bm{v}_{i}(t) +  \bm{f}_{i}(t) \frac{\Delta t}{m_{0}},
\end{equation}
where $\bm{f}_{i}$ is a collision-cell level force with components $f_{\beta} = \frac{a_{0}^{3}}{N_{c}} \partial_{\alpha}(\sigma_{\alpha\beta}^{v,\text{aniso}} + \sigma_{\alpha\beta}^{e})$. Notably the force $\bm{f}_{i}$ represents the backflow effect in our nematic-MPCD model. The divergence of the stress  can be calculated by using again a central difference discretization scheme. With the updated position and velocity of the particles, the collision step can be performed in its original spirit (see App.~\ref{sec:Num_implementation} for the step-by-step algorithm implementation).  

\subsection{Boundary and initial conditions}
One advantage of the MPCD method is the ease with which complicated boundary conditions can be implemented. Here we have tested (a) periodic boundary condition, (b) Lees-Edwards boundary condition, and (c) solid walls. The no-slip condition at solid walls is implemented by using the bounce-back rule with an extra layer of collision cells comprising of virtual particles embedded in the wall \cite{lamura2002numerical, whitmer2010fluid, bolintineanu2012no}. 
A key aspect of the modeling of nematic liquid crystals is the surface anchoring, which not only sets the easy axis for preferred orientation at the surface but also imposes a preferred degree of order. It is very common to have homeotropic (planar) surface anchoring in which the preferred orientation is perpendicular (parallel) to the surface.
In the framework of MPCD, it is convenient to model the surface anchoring using virtual particles \cite{shendruk2015multi}. 

A simple way to model the anchoring would be to assign the tensor order parameter to the virtual particles to the preferred values $Q_{\alpha\beta}^{\text{vp}} = S^{\text{vp}}(3n_{\alpha}^{\text{vp}}n_{\beta}^{\text{vp}} - \delta_{\alpha\beta})/2$ with $S^{\text{vp}}$ as the preferred nematic order and $\bm{n}^{\text{vp}}$ as the preferred orientation at the surface. This choice would impose strong anchoring conditions. Here, however, we adopt a more general way of imposing uniform surface anchoring. Instead of fixing the tensor order parameter of the virtual particles, we update the tensor order parameter of virtual particles as
\begin{equation}\label{Qvp_update_Eq}
\bm{q}_{i}^{\text{vp}}(t + \Delta t) = \bm{q}_{i}^{\text{vp}}(t) + \bm{g}_{i}^{\text{vp}} \Delta t, 
\end{equation} 
where $g_{\alpha\beta}^{\text{vp}} = - (H_{\alpha\beta}^{\text{vp}} + \lambda\delta_{\alpha\beta} + \lambda_{\mu} \epsilon_{\mu\alpha\beta})/\mu_{1}$. Note that we have used a modified molecular field $H_{\alpha\beta}^{\text{vp}}$ for the virtual particles. Incorporating a Rapini--Papoular form of energy term \cite{spencer2006lattice}, $H_{\alpha\beta}^{\text{vp}}$ can be written as
\begin{equation}\label{H_virt_updt}
\begin{split}
H_{\alpha\beta}^{\text{vp}} = & L \partial_{\mu}^{2} Q_{\alpha\beta} - \alpha_{F} Q_{\alpha\beta} + 3\beta_{F} Q_{\alpha\mu}Q_{\beta\mu} \\
& - 4\gamma_{F} Q_{\alpha\beta}Q_{\mu\nu}Q_{\mu\nu} - W^{\text{vp}}(Q_{\alpha\beta} - Q_{\alpha\beta}^{\text{vp}}),
\end{split}
\end{equation}
where $W^{\text{vp}}$ is a phenomenological constant which represents the uniform surface anchoring strength. 
The last term on the right-hand side of Eq.~\eqref{H_virt_updt} is a penalty term which imposes the preferred order and director field. Note that $W^{\text{vp}}$ (in units of N) is a particle-based representation of surface anchoring strength (in units of $\text{N}/\text{m}$) which is commonly used in the liquid crystals literature.        

Initially the MPCD particles are distributed uniformly throughout the simulation domain, while the velocities are taken from the Maxwell--Boltzmann distribution at system temperature $T_{0}$. Depending on the physical situation, we have used two different initial conditions: (a) isotropic state ($i.e.$ $S = 0$), and (b) perfectly nematic state ($i.e.$ $S = 1$).

\subsection{Model parameters}\label{sec:model_parameters}
It is convenient to choose the collision cell length $a_{0}$, the mass of MPCD particle $m_{0}$, and thermal energy $k_{B}T_{0}$ as the scales for length, mass and energy, respectively. 
Scales for other quantities can be derived in the following way: velocity $ v_{0} = \sqrt{k_{B}T_{0}/m_{0}} $, time $t_{0} = a_{0}/v_{0}$, and shear viscosity $\eta_{0} = m_{0}/a_{0}t_{0}$. In our simulations, we consider the collision time step $\Delta t = 0.01 t_{0}$ and mean particle (number) density $\langle\rho \rangle\equiv\langle N_{c} \rangle /V_c= 30 a_{0}^{-3}$, which yields isotropic shear viscosity $\eta_{\text{iso}} = \beta_{4}/2 = 116.274 \eta_{0}$ (this is calculated from MPCD data by performing shear flow simulations \cite{winkler2009stress}). Now, the important task is to map these MPCD units ($a_{0},t_{0}, m_{0}\ \text{and} \ k_{B}T_{0}$) of the coarse-grained system to physical  parameters. 
We consider the common 5CB nematic liquid crystal. Note that the nematodynamic equations of the Qian--Sheng theory have six viscosity coefficients ($\mu_{1}$, $\mu_{2}$, $\beta_{1}$, $\beta_{2}$, $\beta_{4}$, $\beta_{5}$ and $\beta_{6}$), one elastic constant ($L$) and three phenomenological constants ($\alpha_{F}$, $\beta_{F}$ and $\gamma_{F}$). 

\begin{figure}
    \centering
    \includegraphics[width=0.9\columnwidth]{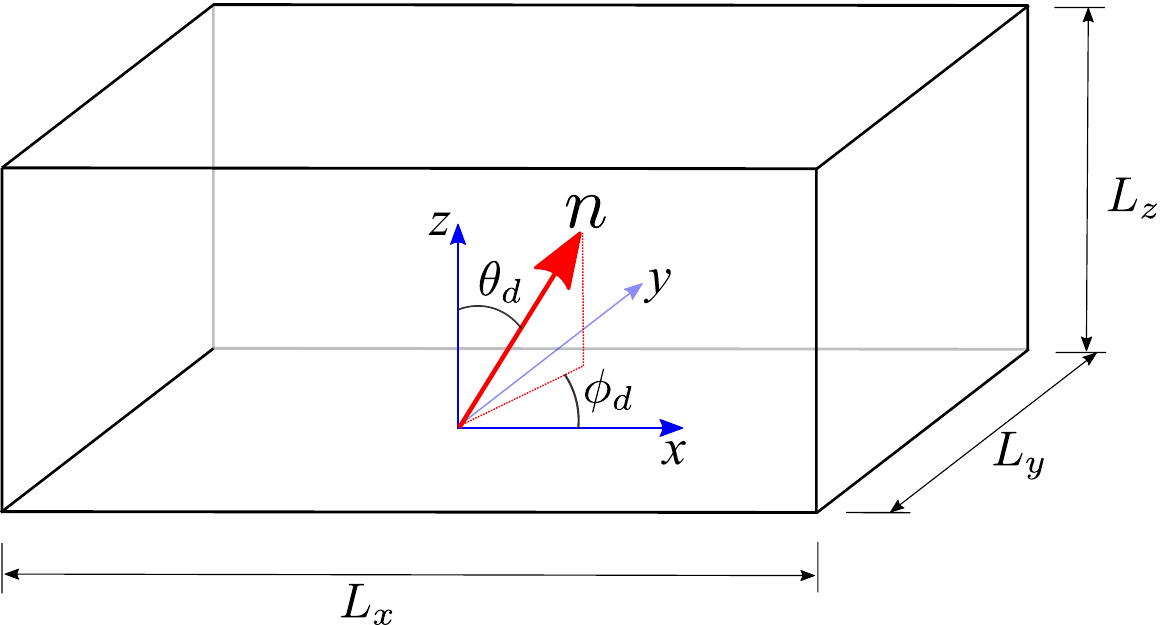}
    \caption{Schematic representation of the simulation setup. The domain is a cuboid box of side lengths $L_{x}$, $L_{y}$, and $L_{z}$ in the $x$, $y$, and $z$ directions of the Cartesian coordinate system. The director $\bm{n}$ is described by two angles $\theta_{d}$ (angle between the director and the $z$ axis) and $\phi_{d}$ (angle between the projection of director on $xy$ plane and $x$ axis).}
    \label{fig:simulation_setup}
\end{figure}

In the limit of constant scalar order parameter,  all  material properties can be expressed in terms of the material properties of the ELP theory in the following way \cite{qian1998generalized} $\mu_{1} = 2(\alpha_{3} - \alpha_{2})/9S_{eq}^2$, $\mu_{2} = 2(\alpha_{2} + \alpha_{3})/3S_{eq}$, $\beta_{1} = 4 \alpha_{1}/9S_{eq}^{2}$, $\beta_{4} = \alpha_{4} + (\alpha_{5} + \alpha_{6})/3$, $\beta_{5} = 2\alpha_{5}/3S_{eq}$, $\beta_{6} = 2\alpha_{6}/3S_{eq}$, and  $L = 2K/9S_{eq}^{2}$, where $S_{eq}$ is the scalar order parameter at equilibrium, $\alpha_{1}$, $\alpha_{2}$, $\alpha_{3}$, $\alpha_{4}$, $\alpha_{5}$ and $\alpha_{6}$ are Leslie viscosities, and $K$ is the Frank elastic constant. Note that the parameters present in the ELP theory can be measured experimentally. Thus, our interest is to find the simulation parameters which will represent a physical system of 5CB nematic liquid crystal (near 26$\celsius$) for which the material properties are \cite{stewart2004static} $\alpha_{1} = -0.0060 \ \text{Pas}$, $\alpha_{2} = -0.0812 \ \text{Pas}$, $\alpha_{3} = -0.0036 \ \text{Pas}$, $\alpha_{4} = 0.0652 \ \text{Pas}$, $\alpha_{5} = 0.0640 \ \text{Pas}$, $\alpha_{6} = -0.0208 \ \text{Pas}$ and $K \sim 6 \times 10^{-12} \ \text{N}$. By performing a mapping, we can determine the  parameters of the Qian--Sheng theory for 5CB near 26$\celsius$ (see App.~\ref{sec:MPCD_mapping} for details).


It is important that our mesoscopic simulations correctly recover the  hydrodynamic state of the physical system. The most important hydrodynamic dimensionless number in the context of nematic liquid crystals is the  Ericksen number $Er = \eta_{\text{iso}} V_{0} L_{0} / K$, where $V_{0}$ and $L_{0}$ are typical velocity and length scales associated with the problem. The Ericksen number signifies the importance of viscous stress relative to elastic stress. Other relevant dimensionless numbers are the Reynolds number $Re$, the Schmidt number $Sc$ and the Mach number $Ma$. To model an incompressible fluid in the Stokes flow regime, we choose  simulation parameters such that $Sc \gg 1$, $Ma < 0.2$, and $Re < 1$.

\section{Results}
The primary goal of this study is to propose a nematic-MPCD method which solves the nematodynamic equations of the Qian--Sheng theory. 
To check the applicability of our nematic-MPCD method, we study several equilibrium and nonequilibrium systems and validate our simulation results with existing results. 
All the simulations are performed in a cuboid simulation domain of size $L_{x} \times L_{y} \times L_{z}$ as depicted in Fig. \ref{fig:simulation_setup}.

\subsection{Nematic-isotropic phase transition}
\begin{figure}
    \centering
    \includegraphics[width=0.45\textwidth]{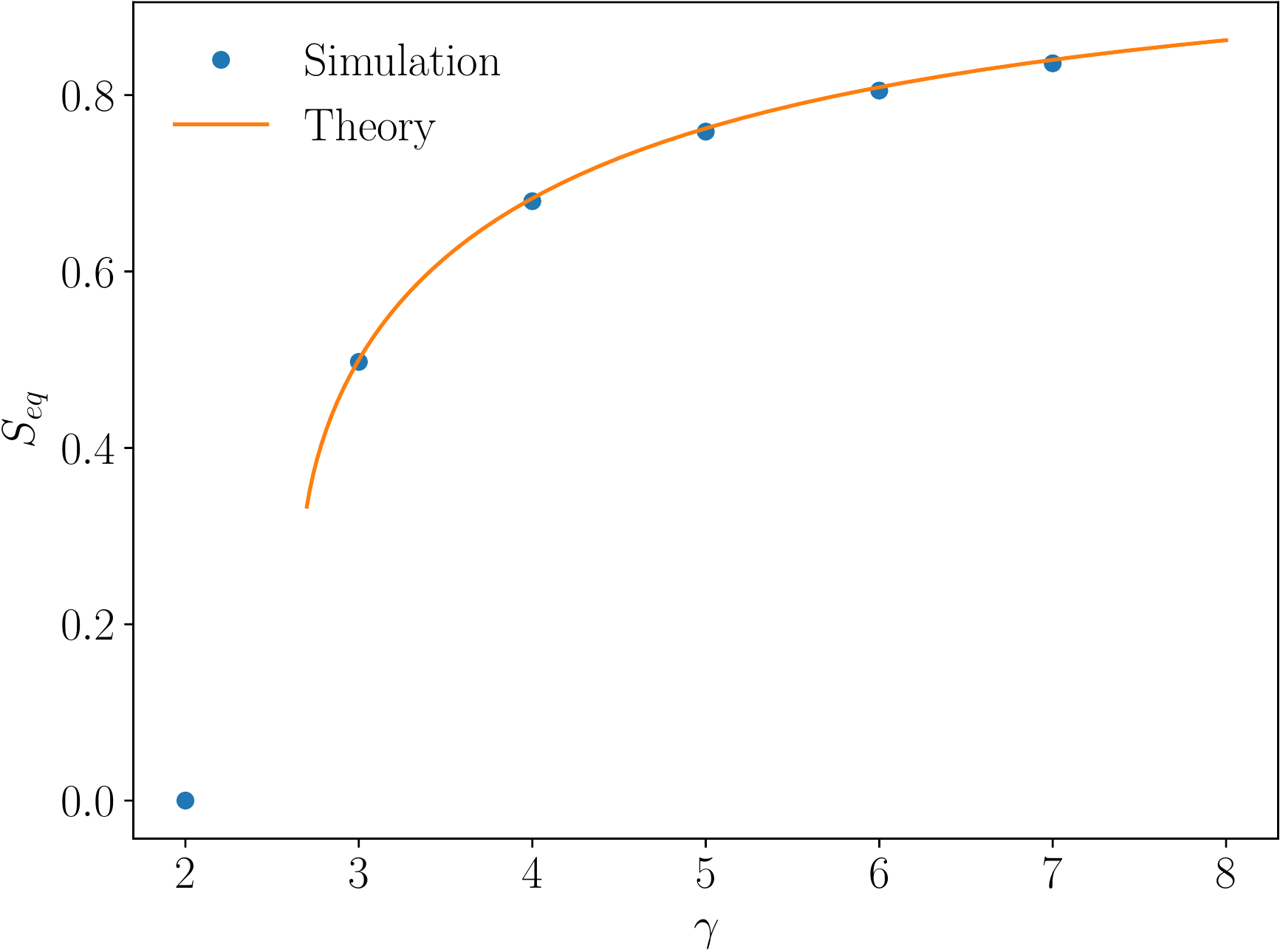}
    \caption{Nematic-isotropic phase transition. Simulations are performed in a 3D box of size $L_{x}=L_{y}=L_{z}=50a_{0}$ with periodic boundary conditions.}
    \label{fig:phase_transition}
\end{figure}

Nematic liquid crystals exhibit a temperature-driven (or concentration driven) first-order phase transition from the ordered nematic phase to a disordered isotropic phase. 
To recover the nematic-isotropic phase diagram, we recast the phenomenological constants present in the Landau--de Gennes free energy as \cite{denniston2001lattice} $\alpha_{F} = A_{0}(1 - \gamma/3)$, $\beta_{F} = 2 A_{0} \gamma/9$, and $\gamma_{F} = 4 A_{0}\gamma/36$, where $A_{0}$ is a constant and $\gamma$ is a parameter which determined the order of the fluid. For thermotropic liquid crystals $\gamma$ represents the effective temperature, while for lyotropic liquid crystals $\gamma$ represents the concentration \cite{lintuvuori2010colloids}. 
By minimizing the Landau--de Gennes free energy, the equilibrium order parameter can be obtained as $S_{eq} = \frac{1}{4} + \frac{3}{4} \sqrt{1 - \frac{8}{3\gamma}}$ in the nematic phase, while $S_{eq}=0$ in the isotropic phase. To reproduce the phase diagram, we have performed simulations in a box of size $L_{x}=L_{y}=L_{z}=50 a_{0}$ with periodic boundary conditions in all three directions. 
We consider $A_{0}=10^{-6} \text{J}/\text{m}^{3}$ and perform simulation for different values of $\gamma$. 
We initialize the system in the isotropic phase and equilibrate the system. 
Upon increasing $\gamma$, the system transitions discontinuously from an isotropic phase at  small $\gamma$ to a nematic phase for large $\gamma$.
Figure \ref{fig:phase_transition} shows an excellent agreement between our simulations and analytical results.

\begin{figure}
    \centering
    \includegraphics[width=0.45\textwidth]{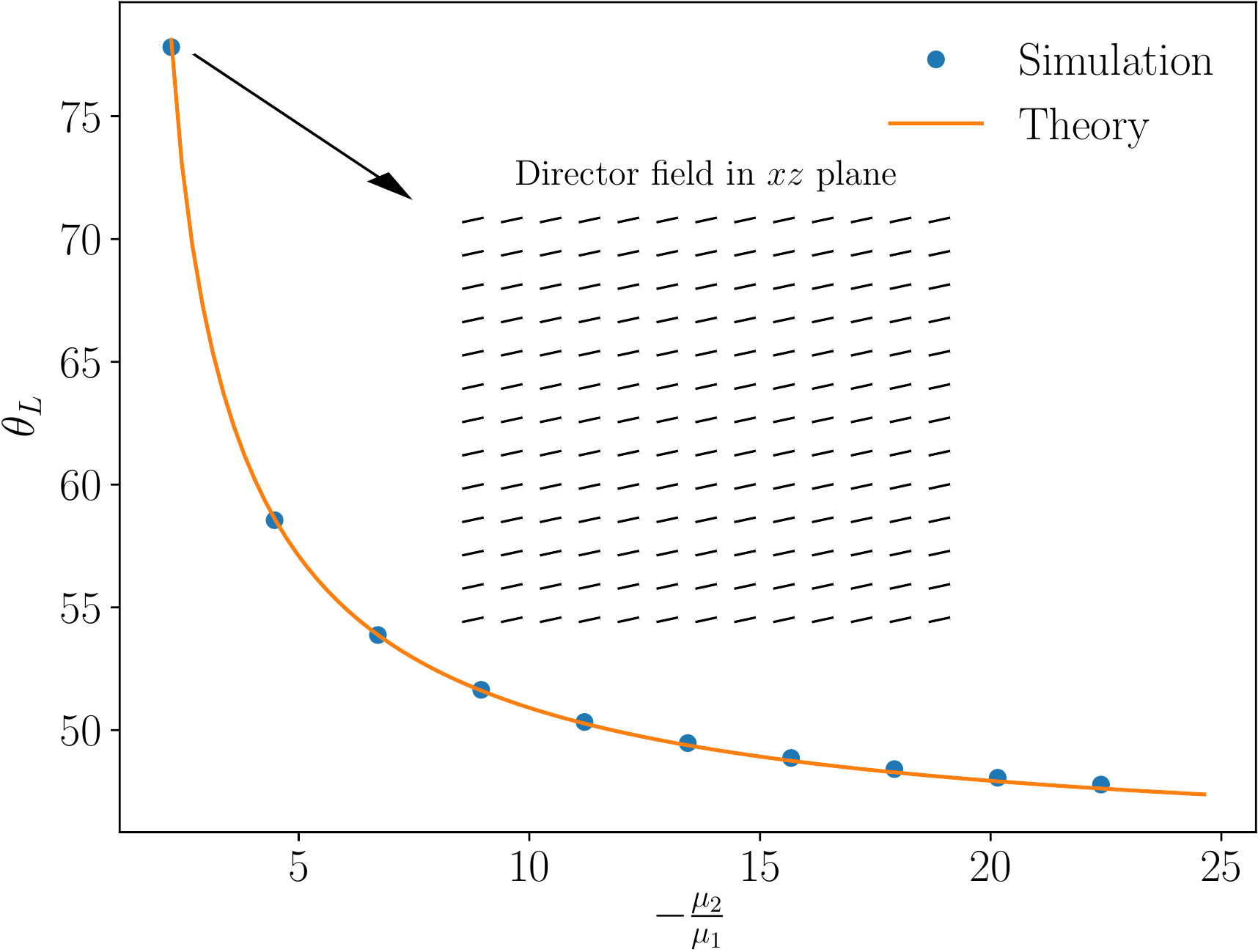}
    \caption{Variation of the Leslie angle $(\theta_{L})$ with the viscosity ratio $(-\mu_{2}/\mu_{1})$ in unbounded shear flow. The inset shows the director (small black dashes) field in $xz$ plane. Simulations are performed in a 3D box of size $L_{x}=L_{y}=L_{z}=50a_{0}$. Periodic boundary conditions are used in $x$ and $y$ directions, while the shear flow is generated by using  Lees--Edwards boundary conditions in the $z$ direction with a shear rate of $0.002t_{0}^{-1}$.}
    \label{fig:shear_alignment}
\end{figure}

\subsection{Director alignment in shear}
Application of a shear flow not only modifies the the scalar order parameter but also reorients the director field. 
Analytical study shows that application of shear flow leads to the alignment of the director field at a particular angle with the flow direction. The director orientation in an unbounded shear flow is solely characterized by the Leslie angle $\theta_{L}$, determined by \cite{spencer2006lattice} $\cos(\pi - 2\theta_{L}) = - 3 S \mu_{1}/\mu_{2}$. 
Depending on the value of viscosity ratio $-\mu_{2}/\mu_{1}$, different director configurations can be obtained: (a) steady flow aligning state, (b) tumbling state, and (c) log-rolling state. In this study, we only focus on the flow-aligning state and perform simulations in a box of size $L_{x}=L_{y}=L_{z}=50a_{0}$ with the Lees--Edwards boundary conditions in the $z$ direction and periodic boundary condition in the $x$ and $y$ directions. 
We initialize the system at a perfectly nematic state with the director field along the direction of shear gradient ($i.e.$ along $z$ direction). Although the flow-orientation coupling is present, the backflow effect is not included here. 
This will allow us to investigate the sole effect of flow coupling on the director orientation. In this limit, the director orientation is solely governed by the viscous torque. 
Figure \ref{fig:shear_alignment} shows the variation of Leslie angle with the viscosity ratio in the flow aligning regime. Upon increasing the viscosity ratio $(-\mu_{2}/\mu_{1})$, the director field tends to align with the principle strain axis $\theta_{L} = 45\degree$. 
The inset shows that irrespective of the position, all the directors align at the same angle $\theta_{L}$ with the $z$  direction. Our simulations compare well with the analytical solution.

\subsection{Shear flow}
In a wall-bounded shear flow, the director orientation is governed by the competition between viscous and elastic torques. 
The elastic torque arises due to the boundary-induced deformation of the director field. This leads to spatial variation of the director orientation in the direction of the shear gradient. 
Following the ELP theory \cite{burghardt1990transient}, a simple one-dimensional analytical model shows that backflow  can significantly modify the linear shear flow profile of a flow-aligning nematic liquid crystal. 
To simulate the combined effect of flow coupling and backflow in shear flow, we consider the nematic liquid crystal between two solid walls (parallel to the $xy$ plane) positioned at $z=0$ and $z=L_{z}$. 
The simulations are performed in a quasi-2D box of size $L_{x}=5L_{y}=L_{z}=50a_{0}$. Instead of using the Lees--Edwards boundary conditions, we use no-slip conditions at both walls. 
Both walls are moving at constant speed $V_{w}$ but in opposite directions. Strong homeotropic anchoring conditions are used at both the walls, while we employ periodic boundary conditions in both $x$ and $y$ directions. We initialize the system in a perfectly nematic state with the director parallel to the $z$ direction. 

\begin{figure}
    \centering
    \includegraphics[width=0.45\textwidth]{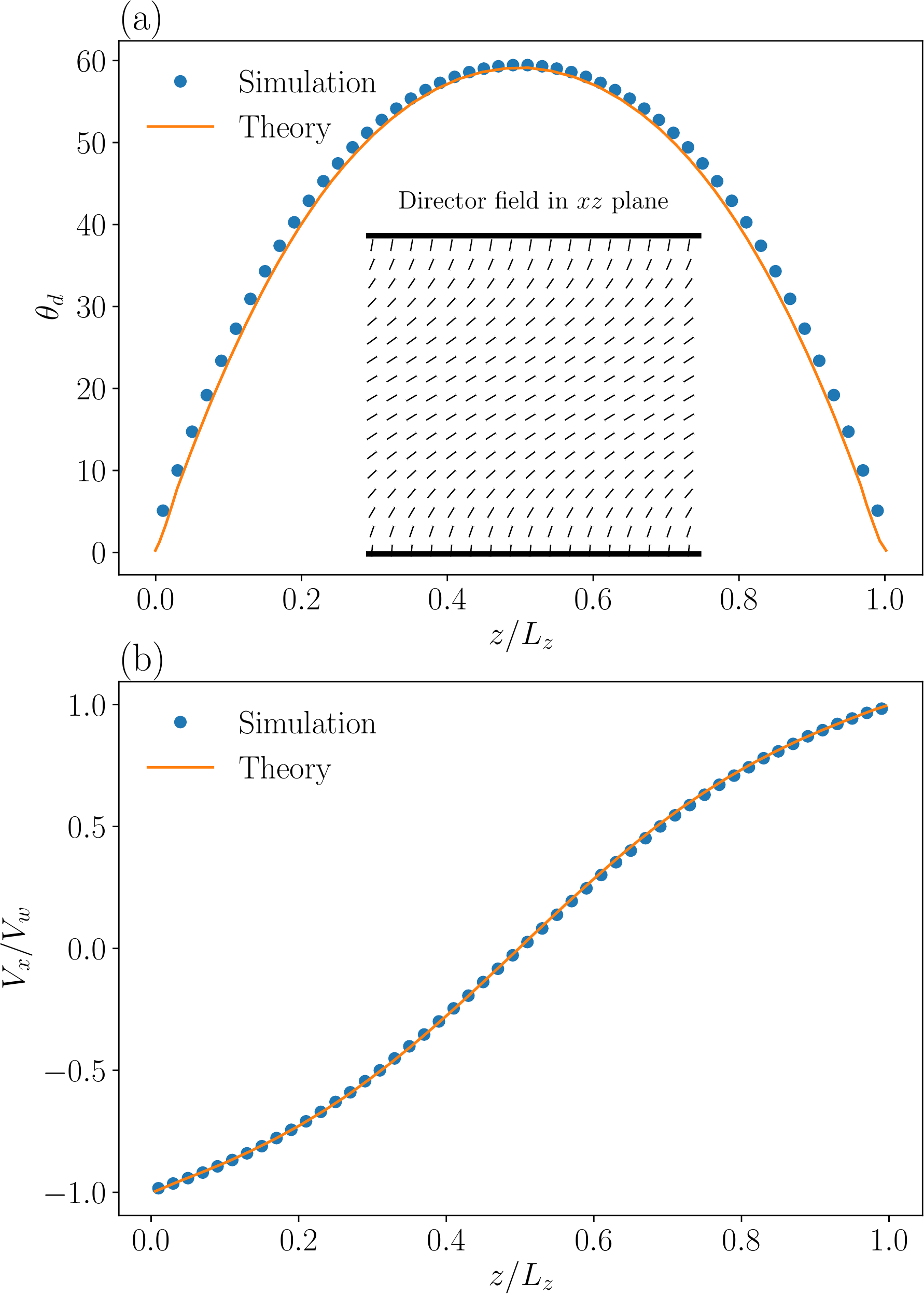}
    \caption{(a) Variation of director orientation angle with the channel height for $Er=5$ in a shear flow. The inset shows the director field in $xz$ plane. (b) Variation of velocity in the shear direction with the channel height for $Er=5$.}
    \label{fig:shear_flow_profiles}
\end{figure}

Figure \ref{fig:shear_flow_profiles}(a) depicts the director profile $\theta_{d}$ for $Er=5$ (for shear flow  $Er \equiv \eta_{\text{iso}} V_{w} L_{z}/K$), and the inset shows the director field (small black dashes). It is evident  that the director remains nearly vertical ($i.e.$ $ \theta_{d} \approx 0 $) near the walls, which is due to strong homeotropic anchoring conditions. Away from the walls, the directors are tilted towards the flow direction, which is due to the velocity-orientation coupling of flow aligning nematics. 
Importantly, the maximum director angle $\theta_{d}(z/L_{z}=0.5)$ is smaller than the Leslie angle $\theta_{L} \approx 78\degree$. 
This reflects the fact that the director orientation in a bounded domain is determined by the combined action of viscous and elastic torques. 

The velocity profile is shown in Fig. \ref{fig:shear_flow_profiles}(b). In sharp contrast to the velocity profile of an isotropic fluid, the velocity profile of nematic fluid deviates significantly from a linear profile. Note that the velocity gradient across the channel height is not constant any more. 
The velocity gradient is larger near the center as compared to the velocity gradient near walls. This is due to the backflow effect present in nematic liquid crystals. The director field near the center is aligned more towards the flow direction as compared to the director field near the walls. This orientation pattern reduces the effective viscosity near the center and leads to increase in velocity gradient. 
For both director profile and velocity profile we have obtained excellent agreement with the analytical solutions as depicted in Fig. \ref{fig:shear_flow_profiles}.          

\begin{figure}
    \centering
    \includegraphics[width=0.45\textwidth]{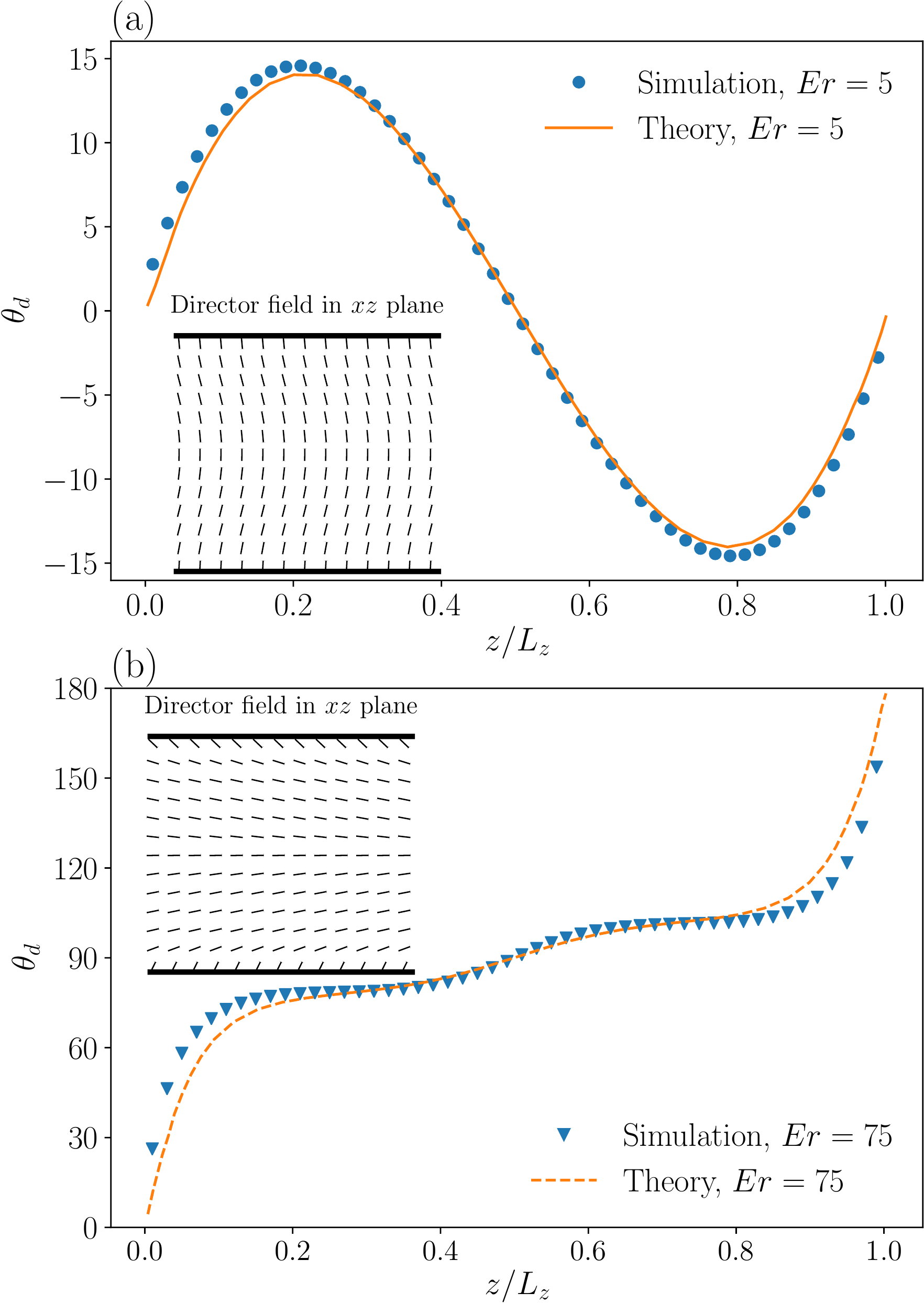}
    \caption{Variation of the director orientation angle with the channel height in a Poiseuille flow for (a) V-state, and (b) H-state. The insets show the director field in $xz$ plane. }
    \label{fig:director_ang_poiseuille_flow}
\end{figure}
\begin{figure}
    \centering
    \includegraphics[width=0.45\textwidth]{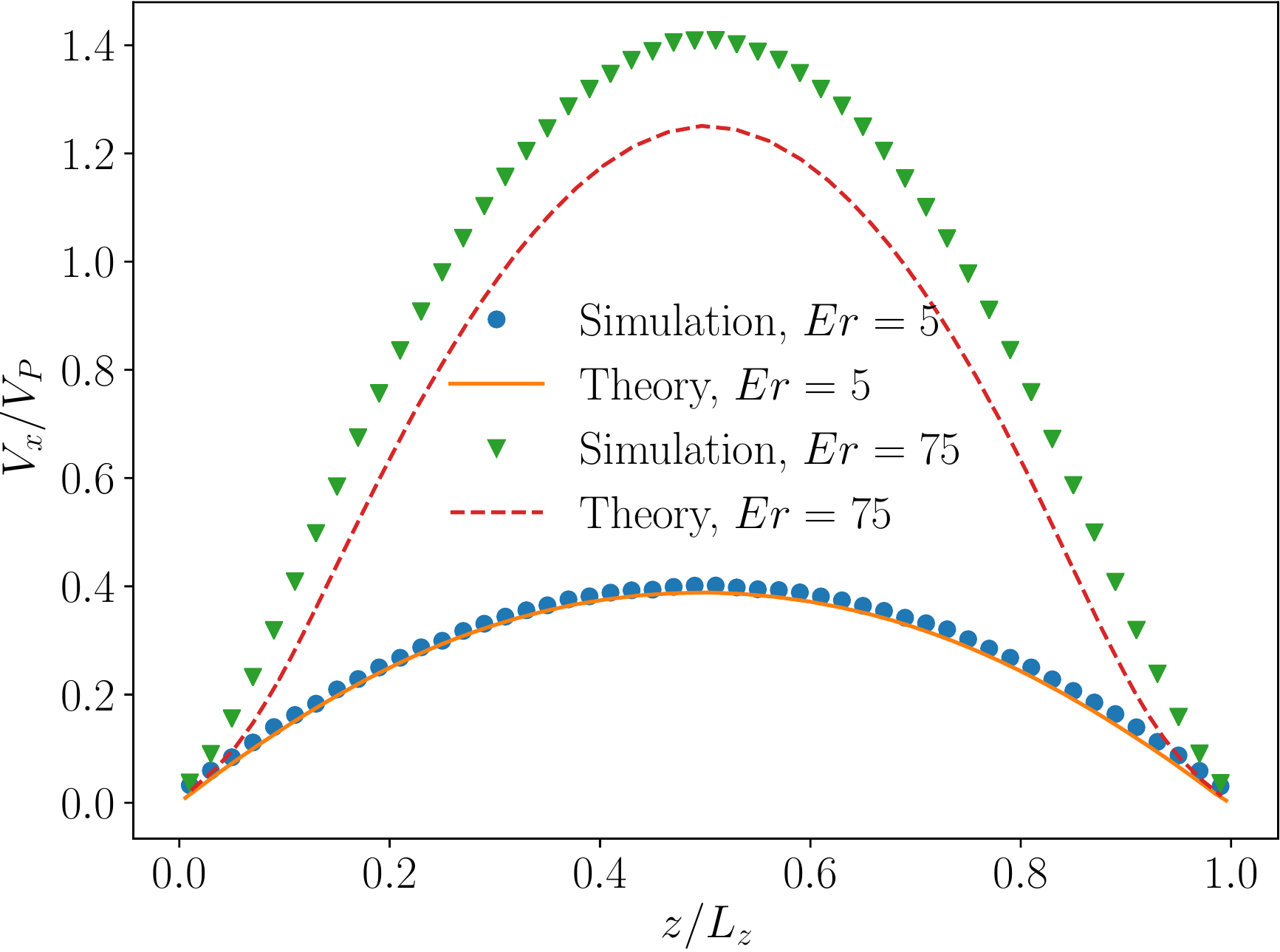}
    \caption{Variation of the velocity in the direction of the Poiseuille flow with the channel height for a V-state ($Er=5$), and  H-state ($Er=75$).}
    \label{fig:velocity_poiseuille_flow}
\end{figure}

\subsection{Poiseuille flow}
We turn next to the Poiseuille flow of flow-aligning nematic liquid crystals. The simulation domain is similar to the shear flow setup but with two stationary solid walls at $z=0$ and $z=L_{z}$, and flow along the $x$ axis. Strong homeotropic surface anchoring conditions are used for both the walls. 
The initial director orientation is set perpendicular to the walls. The Poiseuille flow can be induced by applying on each MPCD particle a constant force $f_{P}$ in the flow direction. Experiments \cite{jewell2009flow, sengupta2013liquid} have shown that the director field can attain topologically distinct profiles in a Poiseuille flow depending on the volumetric flow rate. 
Recently, simulation studies \cite{denniston2001simulations, batista2015effect} have also reported similar results. To investigate this, we perform simulations for different $Er$. Here $Er$ is defined as $Er = \eta_{\text{iso}}V_{P}L_{z}/K$, where $V_{P} = f_{P} \langle N_{c} \rangle L_{z}^{2}/(8\eta_{\text{iso}}a_{0}^{3})$ is the center-line velocity for an isotropic fluid. 

For small $Er$ ($i.e.$ low flow regime), we obtain a stable director configuration in which the director aligns perpendicular to the flow at the channel center, whereas for large $Er$ ($i.e.$ high flow regime) we obtain a different steady-state director configuration in which the director aligns parallel to the flow at the channel center. Figures \ref{fig:director_ang_poiseuille_flow}(a) and (b) depict these director profiles and director fields for $Er=5$ and $Er=75$, respectively. 
These two configurations are called vertical (V) and horizontal (H) states, respectively. Note that the bend deformation is more prominent for V-state, while the splay deformation is more prominent for H-state. 
A closer look into the temporal evolution of director field reveals that the transition from the V-state to the H-state at high-flow regime takes place through appearance and subsequent disappearance of a topological defect at the channel center. Figure \ref{fig:director_ang_poiseuille_flow} shows that our simulation compares well with the theory \cite{anderson2015transitions, jewell2009flow} for small $Er$, while for large $Er$ the the near-wall directors obtained from simulation are more aligned towards the flow direction as compared to the theoretically obtained directors. 
This is due to the fact that the theory is derived for infinitely strong anchoring, while our simulations are  performed using a more general form of anchoring condition which takes into account the effect of elastic deformation and flow field on the director field at the wall. 
Note that the near the wall viscous and elastic forces can change the director orientation away from the preferential value ($i.e.$ anchoring condition) which is not captured in the theory.

The velocity profiles are also very different in these two configurations as depicted in Fig. \ref{fig:velocity_poiseuille_flow}. For small $Er$, the elastic stress resists the flow and leads to a reduction of flow velocity at the center-line as compared to an isotropic fluid (represented by $V_{x}/V_{P} < 1$). On the other hand, for $Er=75$ the flow velocity at the center-line is significantly larger as compared to the isotropic fluid ($V_{x}/V_{P} > 1$). 
This is due to the fact that the director configuration parallel to the flow direction associated with the H-state reduces the effective viscosity. Figure \ref{fig:velocity_poiseuille_flow} shows that our simulations compare well with the theory for small $Er$, while for large $Er$ the simulation shows larger velocity due to the effects of near-wall directors which are aligned more towards the flow direction.   

Note that the director profile in Poiseuille flow is also  strongly  dependent on the initial state  \cite{denniston2001simulations}. Our simulations show that the system can exhibit the H-state even for small $Er$ provided the initial state of the fluid is isotropic and the system is quenched afterwards.

\subsection{Defect annihilation}
\begin{figure}
    \centering
    \includegraphics[width=0.45\textwidth]{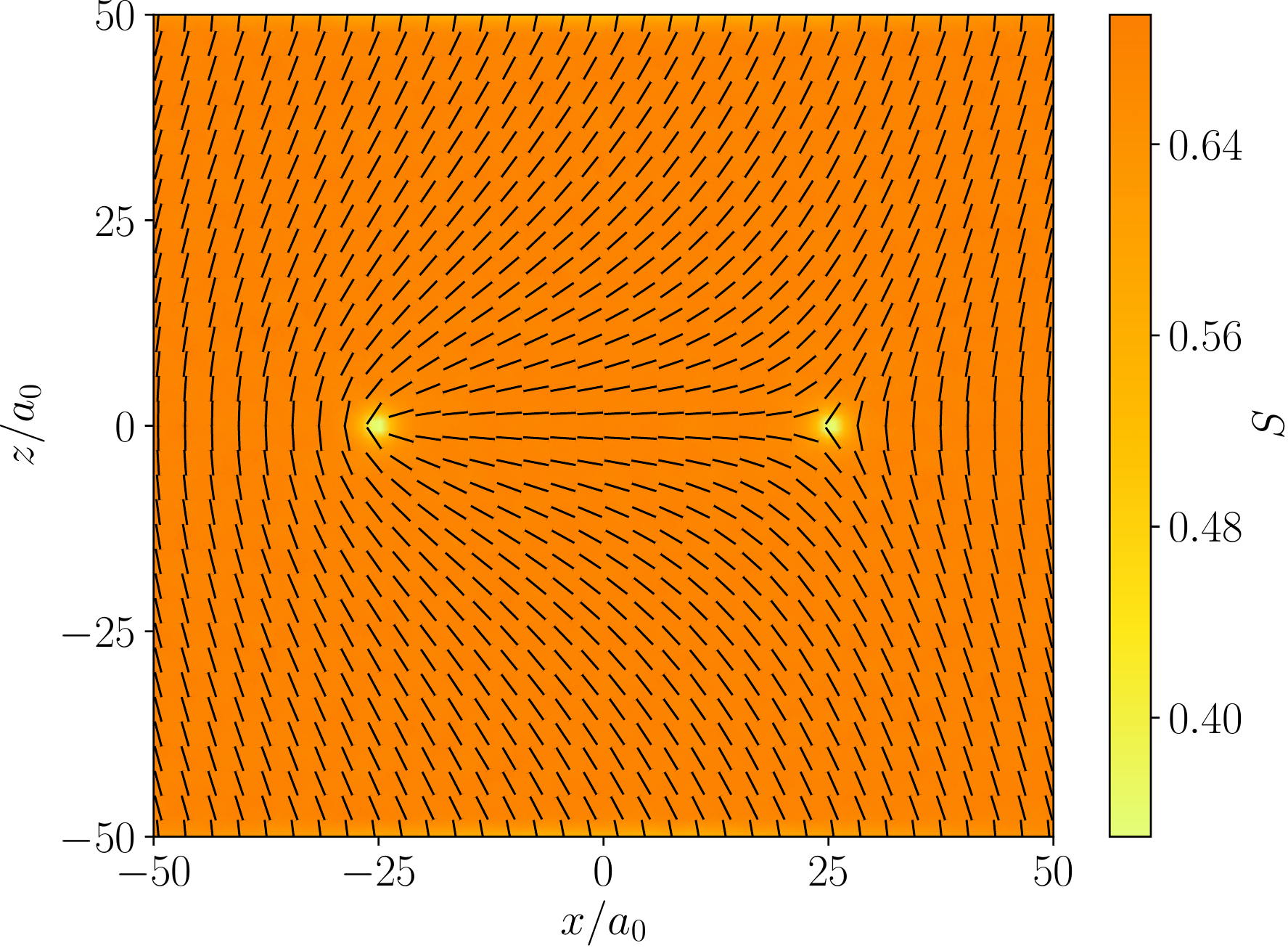}
    \caption{Initial configuration of a pair of $\pm 1/2$ topological defects. The small dashes represent the director field, while the color shows the scalar order parameter. The nematic order parameter is averaged over the $y$ direction.}
    \label{fig:director_field_annihilation}
\end{figure}

Topological defects are regions where the director field is not defined and with a very small value of scalar order parameter. Defects might arise due to quenching of the system from isotropic to nematic state, imposition of surface anchoring conditions or application of external fields \cite{lavrentovich2012defects}. 
A description based on variable order parameter is of great importance to study the defect dynamics as the components of $\bm{Q}$ are continuous within a defect core whereas the director field $\bm{n}$ is often discontinuous. 
As a test case,  we simulate the annihilation dynamics of two $\pm 1/2$ line defects.
We consider a quasi-2D domain of size $L_{x}=10L_{y}=L_{z}=100a_{0}$ with periodic boundary condition in all three directions. The system is initialized with a predefined tensor order parameter as $Q_{\alpha\beta} = (3 n_{\alpha} n_{\beta} - \delta_{\alpha\beta})/2$, where $\bm{n} = \sin\theta \bm{e}_{x} + \cos\theta \bm{e}_{z}$ and $\theta = \left[ \frac{1}{2}  \tan^{-1} \left( \frac{z-z_{p}}{x-x_{p}} \right) - \frac{1}{2} \tan^{-1} \left( \frac{z-z_{m}}{x-x_{m}} \right) \right] $. This initialization places $+1/2$ defect line and $-1/2$ defect line at locations $(x_{p},z_{p})$ and $(x_{m}, z_{m})$ as depicted in Fig. \ref{fig:director_field_annihilation}. Here we place the two defect lines at an initial separation distance of $50 a_{0}$ along the $x$ direction.   

\begin{figure}
    \centering
    \includegraphics[width=0.45\textwidth]{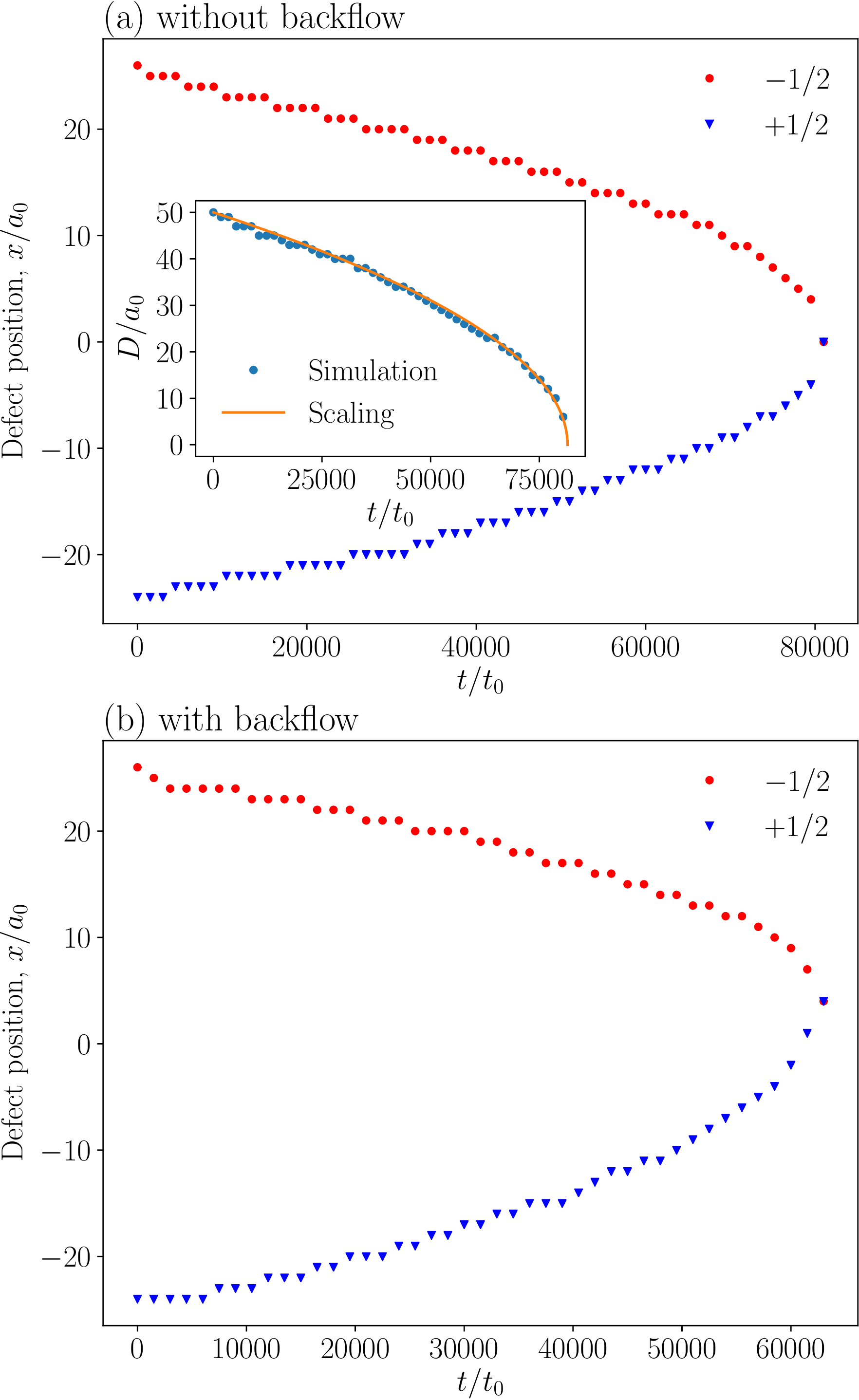}
    \caption{Temporal evolution of the defect positions (a) without backflow, and (b) with backflow. The inset shows the temporal evolution of the distance between the two defects.}
    \label{fig:annihilation_defect_pos}
\end{figure}

First, we study the annihilation dynamics without the backflow effect. Figure \ref{fig:annihilation_defect_pos}(a) depicts the temporal evolution of the defects' positions with time (the defect location is determined by calculating the position of the minimum of the scalar order parameter). Both defects move towards each other, initially with a small speed. 
As time progresses, both  defects accelerate, and eventually meet and annihilate. This defect motion is solely driven by the molecular field which always drives the system to minimize its free energy. 
The motion of both defects is symmetric as they  move with equal speed and they meet at the midpoint of their initial separation. Similar results have also been reported before \cite{svenvsek2002hydrodynamics, toth2002hydrodynamics}. A simple analytical model shows that the separation distance between the two defects $D$ follows a scaling law of the form \cite{denniston1996disclination} $D(t) = c\sqrt{t_{a} - t}$, where $t_{a}$ is the annihilation time and $c$ is a constant. 
Our simulation results compare well with this scaling behaviour as shown in the inset of Fig. \ref{fig:annihilation_defect_pos}(a). 
Backflow significantly alters the annihilation dynamics [Fig. \ref{fig:annihilation_defect_pos}(b)]. The annihilation process is much faster as compared to the no-backflow case. Also, the speed of $+1/2$ defect is considerably larger than the speed of $-1/2$ defect. This leads to the asymmetric motion of the defects. The initial, deformed director field generates elastic stresses which lead to the generation of hydrodynamic stresses. This hydrodynamic flow created by the backflow mechanism significantly affects the annihilation dynamics as also reported in previous studies \cite{svenvsek2002hydrodynamics, toth2002hydrodynamics}.    

\subsection{Force dipole}
\begin{figure}
    \centering
    \includegraphics[width=0.6\columnwidth]{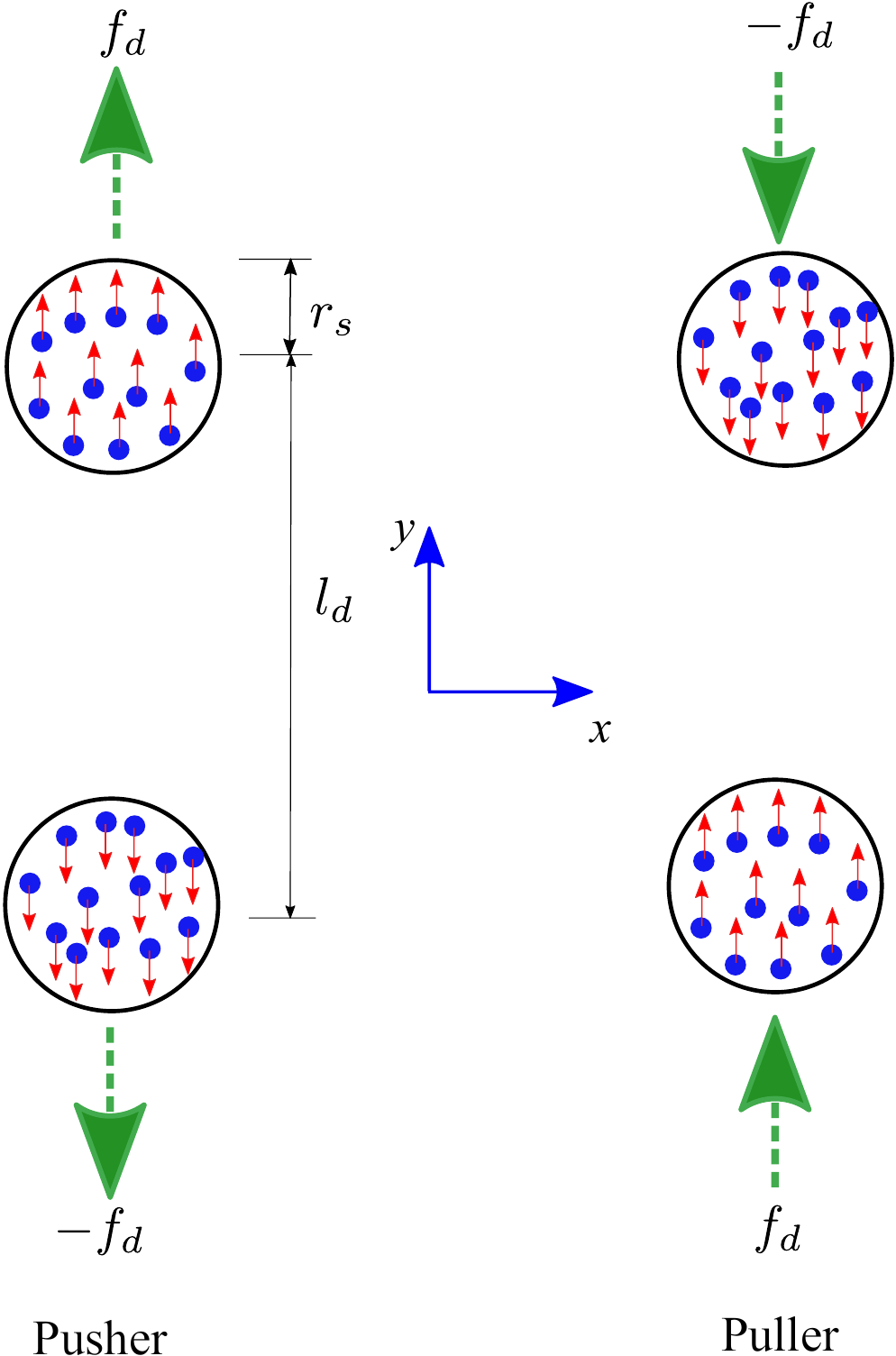}
    \caption{Schematic representation of the force dipole for pusher-type (left) and puller-type (right) in our particle-based framework. The macroscopic force of magnitude $f_{d}$ is distributed among the MPCD particles which are present at that instant of time. The blue dots represent the MPCD particles. The green arrows represent the macroscopic force, while the red arrows represent the contribution to the velocity of MPCD particles due to application of the two Stokeslets.}
    \label{fig:force_dipole_schematic}
\end{figure}

We now turn to some new aspects of nematodynamics, namely the effect of force (Stokeslet) dipole on a nematic fluid. It is well known that the leading-order flow field created by several biological microswimmers ($e.g.$ \textit{Escherichia coli} and \textit{Chlamydomonas}) can be modeled as a force dipole \cite{drescherPRL2010, drescherPNAS2011, elgeti2015physics, lauga2009hydrodynamics}. 
Pusher- (puller-) type force dipoles represent the leading-order flow field of \textit{Escherichia coli} (\textit{Chlamydomonas}) \cite{baskaran2009statistical}. Very recently, Kos and Ravnik \cite{kos2018elementary} studied analytically the flow field around a force dipole in nematic liquid crystal assuming spatially uniform director field. 
However, there are physical situations in which the presence of microswimmer deforms the director field not only due to the surface anchoring condition, but also due to the velocity-orientation coupling \cite{sokolov2015individual, zhou2017dynamic}. 
Here, we perform simulations which naturally account for the velocity-orientation coupling. To implement a regularized force dipole, we identify two spherical regions of size $r_{s}=1.5a_{0}$, which are $l_{d}=10a_{0}$ distance apart on the $xy$ plane. This uniaxial force configuration along the symmetry axis of the force dipole generates a stresslet flow field. The macroscopic point force of magnitude $f_{d} = 500 k_{B}T_{0}/a_{0}$ is distributed among the MPCD particles within the spherical regions as depicted in Fig. \ref{fig:force_dipole_schematic}. 
A similar MPCD technique was recently used to model microswimmers \cite{schwarzendahl2018maximum}. We consider a simulation box of size $L_{x}=L_{y}=L_{z}=50a_{0}$, and apply strong planar anchoring at the two solid walls (located at $z=0$ and $z=L_{z}$). 

\begin{figure*}
    \includegraphics[width=0.8\textwidth]{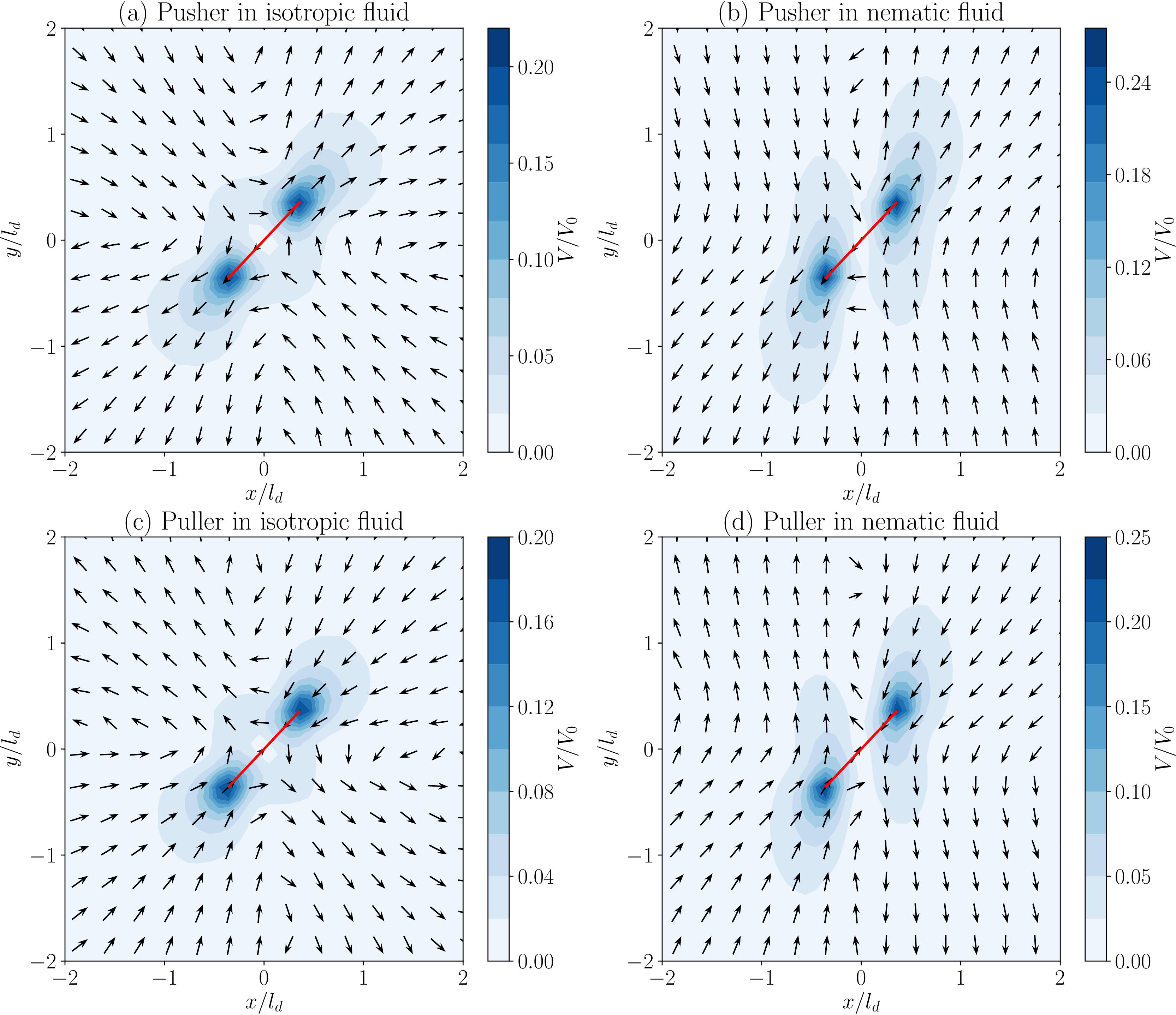}
    \caption{Time-averaged flow field generated by: (a) pusher-type force dipole in an isotropic fluid, (b) pusher-type force dipole in a nematic fluid, (c) puller-type force dipole in an isotropic fluid, and (d) puller-type force dipole in a nematic fluid. The vectors represent the direction of flow, while the color shading represents the magnitude of velocity. The force dipole is placed in $xy$ plane. The dipole axis (shown by red solid line) makes an angle $\pi/4$ with the $x$ direction. The director field in the bulk is along $y$ direction.}
    \label{fig:dipole_flow_field}
\end{figure*}

Very recent studies have reported that at steady state a pusher-type microswimmer swims along the director field, while puller-types  swim perpendicular to it \cite{lintuvuori2017hydrodynamics, daddi2018dynamics}. To explore a range of possibilities, we simulate a force dipole tilted with respect to the far-field director. We place the force dipole at an angle $\phi = \pi/4$ from the $x$ direction, while the far-field director is aligned along the $y$ direction. Figure \ref{fig:dipole_flow_field} shows the flow field for pusher-type (a,b) and puller-type (c,d) force dipoles in an isotropic (a,c) and nematic (b,d) fluid, respectively. 

Firstly, in an isotropic fluid, the flow field is symmetric about the dipole axis for both pusher- and puller-type force dipoles. But this symmetry is not preserved in the nematic fluid. Second, the anisotropic medium not only affects the magnitude of velocity, but also stretches the velocity field in the direction of director field ($i.e.$ $y$ direction). This is due to the fact that that the resistance to flow is less along the director as compared to the direction perpendicular to director. This leads to a larger component of flow along the director. Third, close inspection of the velocity field in Fig. \ref{fig:dipole_flow_field}(b) reveals that the velocity vectors on the right-hand side are more aligned to the upward direction (positive $y$) and the velocity vectors on the left-hand side are more aligned to the downward direction (negative $y$). 
This biased flow field effectively resembles a rotational flow around the pusher-type force dipole in the counter-clockwise direction. This gives a clear indication about the presence of a hydrodynamic torque experienced by a pusher-type microswimmer. This torque will try to align the pusher-type microswimmer along the director field. 
The opposite situation is observed for the puller-type force dipole in Fig. \ref{fig:dipole_flow_field}(d). The puller-type force dipole will experience a hydrodynamic torque in the clockwise direction which will try to align it perpendicular to the director field. Interestingly, though our simulations do not include the swimmer's solid body, the observations from the flow field around the force dipole can explain the hydrodynamic torque-induced alignment of pushers (pullers) along (perpendicular to) the director as also reported for squirmer motion in nematics  by Lintuvuori, \emph{et al.} \cite{lintuvuori2017hydrodynamics}.      

\begin{figure}
    \includegraphics[width=0.99\columnwidth]{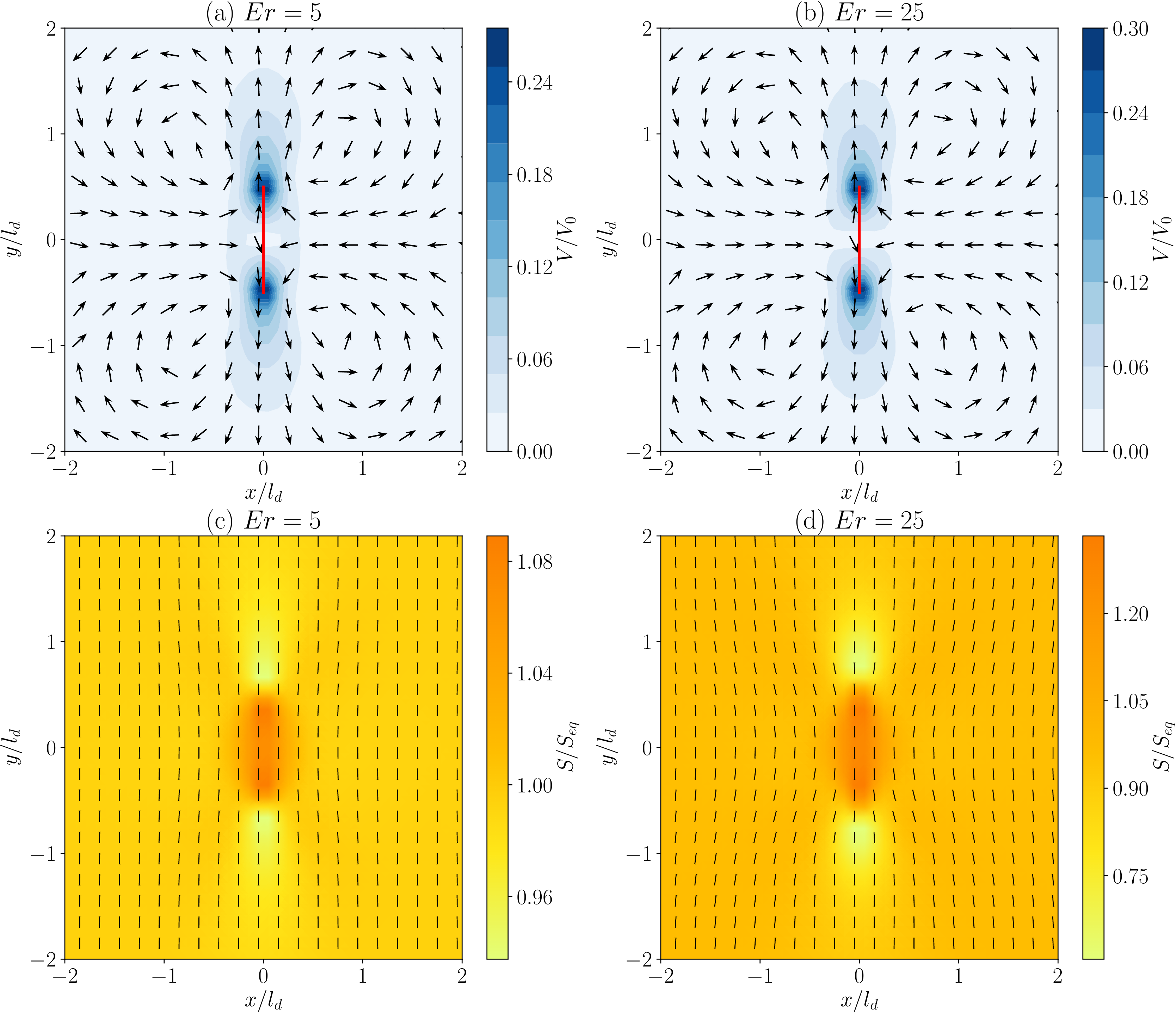}
    \caption{Time-averaged flow field for (a) $Er=5$, and (b) $Er=25$. Time-averaged director field for (c) $Er=5$ and (d) $Er=25$. The pusher-type force  dipole is now placed parallel to the far-field director field.}
    \label{fig:pusher_flow_director_field}
\end{figure}

\begin{figure}
    \includegraphics[width=0.99\columnwidth]{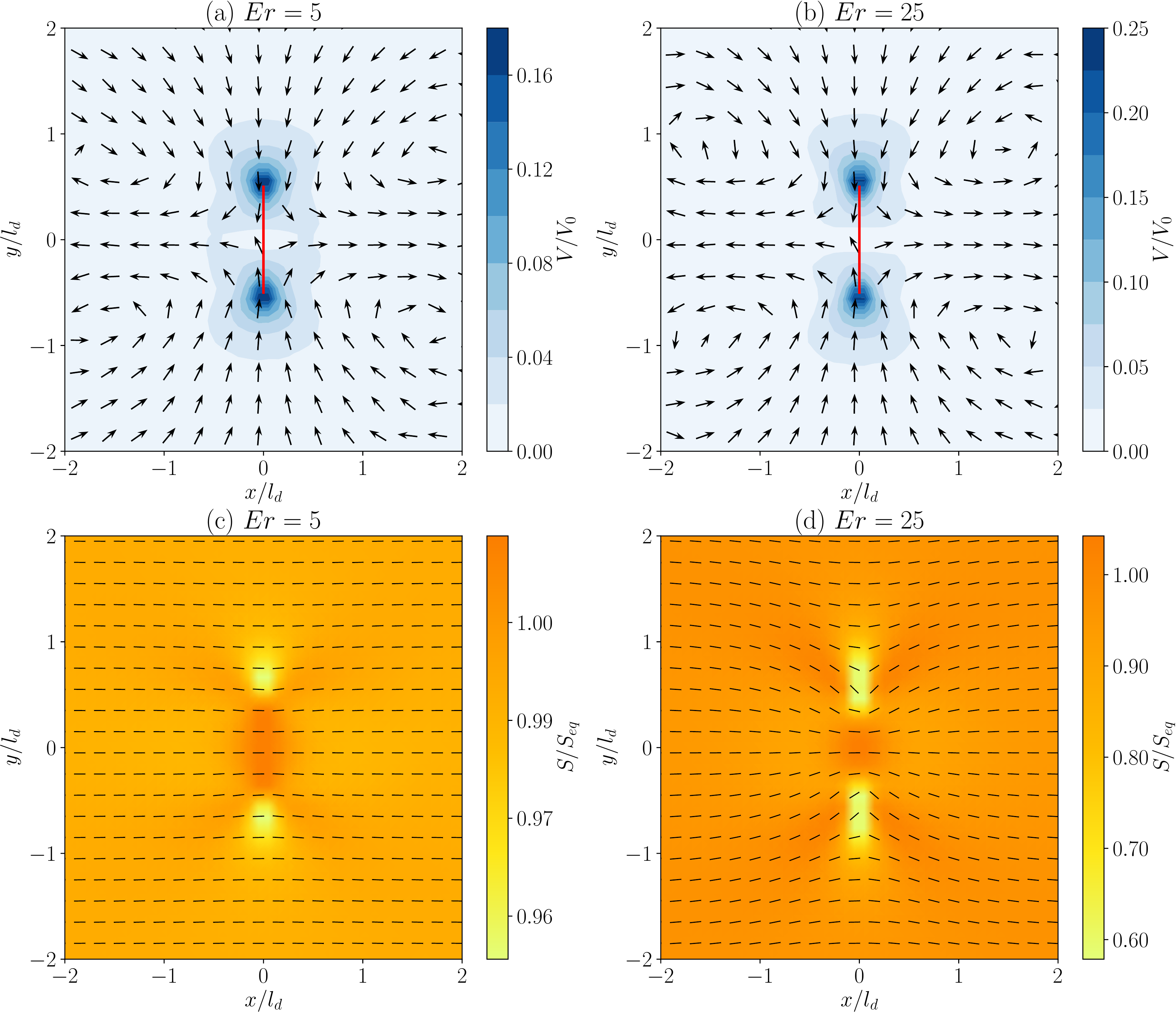}
    \caption{Time-averaged flow field for (a) $Er=5$, and (b) $Er=25$. Time-averaged director field for (c) $Er=5$ and (d) $Er=25$. The puller-type force  dipole is now placed perpendicular to the far-field director field.}
    \label{fig:puller_flow_director_field}
\end{figure}

We now investigate the effects of $Er$ on the velocity field and director field in the `preferred' configuration of the microswimmers, that is, parallel to the director for pushers, and perpendicular to it for pullers. In this context,   $Er \equiv \eta_{\text{iso}}V_{0}l_{d}/K$, where $V_{0} = f_{d}/(\eta_{\text{iso}} l_{d})$. 
Figure \ref{fig:pusher_flow_director_field} depicts the velocity and director fields for a pusher-type force dipole. 
For small $Er$, there is no noticeable change in director field as the directors remains mostly parallel to the dipole axis. For larger $Er$ we observe deformations of the director field around the force dipole. The effect of director deformation is associated to an increase in magnitude of velocity. Note that the director deformation on the $xy$ plane remains symmetric about the dipole axis. 
Figure \ref{fig:puller_flow_director_field} depicts the velocity and director fields for a puller-type force dipole. 
For small value of $Er$, there is no noticeable change in director field as the directors remain mostly perpendicular to the dipole axis. For larger $Er$, we observe deformations of the director field around the force dipole. Similarly to the pusher case, the effect of director deformation is associated to an increase in magnitude of velocity,  and the director deformation is symmetric about the dipole axis. The director deformation is larger for puller-type force dipoles than for pusher-type force dipoles as the strong flow is perpendicular to the director field for puller-type force dipole configuration.

\section{Conclusions}
We have proposed a new mesoscopic simulation technique for fluctuating nematodynamics. 
Following the Qian--Sheng theory of nematodynamics, we have extended the MPCD method to model nematic liquid crystals with variable tensor order parameter. The nematic orientational order is incorporated in the particle-based formulation by assigning a tensor order parameter to each MPCD particle. 
Two-way coupling (velocity-orientation coupling and backflow effect) is duly incorporated by using mesoscopic derivatives. We have described the proposed method in three dimensions. Surface anchoring is also implemented using virtual particles. 

We employed physical  parameters associated to a common nematic liquid crystal (5CB). 
In order to validate our new method, we have simulated the equilibrium nematic-isotropic phase transition behavior, which compares well with the Landau--de Gennes theory. Shear flow-induced alignment of the director is also reproduced by simulating the shear flow using Lees--Edwards boundary conditions. The variation of the director angle with the viscosity ratio compares well with the existing theory. 
We implement shear and Poiseuille flows in a parallel-plate geometry bounded by two walls with homeotropic surface anchoring conditions. The director and velocity profiles compare well with the existing theory. A more stringent test of our implementation of nematodynamics is afforded by the study of how  a pair of $\pm 1/2$ line defects annihilate. The effect of backflow is reproduced. These extensive checks signify the fact that the proposed MPCD method correctly incorporates the flow coupling and backflow mechanisms. 
We have also studied the deformations due to a force dipole embedded in the nematic liquid crystal. The orientation of dipole axis relative to the far-field director field is found to have a pronounced effect on the flow field. 
When the force dipole is tilted with respect to the far-field director, the coupling between the stresslet flow field and the director field generates a rotational component of flow which is reminiscent of a hydrodynamic torque. The flow fields around the pusher- and puller-type force dipoles indicate that a pusher tends to align with the director, while the puller tends to align perpendicular to the director.   

There are numerous avenues in which our proposed novel method can be extended. One advantage of the proposed method as compared to the existing lattice Boltzmann or finite difference/element methods is that our method contains thermal fluctuations. 
Thus the present method can be used to study the dynamics of colloids and microswimmers immersed in nematic liquid crystals. The proposed method, however, is limited to uniaxial nematics. Thus, additional extensions include  biaxial and chiral nematic liquid crystals. External electric/magnetic fields could be easily incorporated in the present formulation.

\section*{Acknowledgements}
SM gratefully acknowledges Joscha Tabet, Dr. Fabian Schwarzendahl,  and J\'er\'emy Vachier for insightful discussions. SM and MGM also gratefully acknowledge support from the the German Research Foundation (DFG) Priority Program SPP1726 \textquotedblleft Microswimmers\textquotedblright.

\appendix

\section{Numerical implementation} \label{sec:Num_implementation}
Here we describe the numerical implementation of the algorithm for a typical simulation system in which the domain is bounded by two stationary solid walls. 
In the presence of bounding walls, the system consists of fluid particles and virtual particles. The fluid particles represent the nematic fluid, while the virtual particles are used to impose no-slip conditions and surface anchoring  at the solid walls. First, we initialize the particles' positions, velocities, and nematic order parameters. The fluid particles are distributed uniformly inside the domain with average number of particles per cell $\langle N_{c}\rangle$.

Solid walls are represented by an extra layer of collision cells. Thus, the thickness of the layer containing the virtual particles inside the solid wall is $a_{0}$.
The virtual particles are also uniformly distributed inside the walls with the same number density.  All the particles are assigned initial velocities sampled from the Maxwell--Boltzmann distribution with $\sqrt{k_{B}T_{0}/m_{0}}$ as standard deviation and zero mean. The initial linear and angular momentum are removed from each cell and the velocities are rescaled as per system temperature $T_{0}$. The nematic order parameter is also initialized as discussed above. At each time step, the following key steps are implemented on CUDA-capable GPU to calculate particle position, velocity and tensor order parameter:

\begin{enumerate}
  \item All the particles (fluid and virtual) are sorted in their respective cells and cell-level quantities ($i.e.$ $N_{c}$, $\bm{Q}_{c}$ and $\bm{V}_{c}$) are calculated. 
  
  \item The mesoscopic derivatives present in $g_{\alpha\beta}$ and $f_{\beta}$ are calculated by using a central difference discretization scheme \cite{eisenstecken2018hydrodynamics}. For any mesoscopic quantity $\psi$, the central difference scheme reads as $\partial_{\alpha}\psi \approx (\psi_{\alpha+1} - \psi_{\alpha-1})/2$. This discretization ensures that the total force acting on the particles due to the divergence of anisotropic viscous stress and elastic stress is zero, and there is no macroscopic momentum drift in the absence of external forces.
  
  \item The particle-based tensor order parameter of fluid particles and virtual particles $\bm{q}_{i}$ and $\bm{q}_{i}^{\text{vp}}$ are updated using Eq.~\eqref{Q_update_Eq} and \eqref{Qvp_update_Eq}, respectively. We have only solved for the 6 independent components of $\bm{q}_{i}$. The symmetry of the tensor order parameter is assumed. 
  
  \item The position and velocity of the fluid particles are updated by performing the streaming step following Eq.~\eqref{Mod_Streaming_v}. Boundary conditions are applied if a particle crosses the system boundary. The bounce-back rule is used when fluid particle collides with a solid wall. The particle velocity is reversed at the point of collision and the particle is moved for the rest of the trajectory with updated velocity $\bm{v}_{i} \leftarrow - \bm{v}_{i}$. Position of the virtual particles $\bm{r}_{i}^{\text{vp}}$ are randomly drawn from a uniform distribution.
  
  \item Galilean invariance is violated due to partitioning of the system into a grid of collision cells \cite{ihle2001stochastic}. To reestablish the Galilean invariance, we move all the particles (keeping collision grid fixed) by a random vector $\bm{s}$ as $\bm{r}_{i} \rightarrow (\bm{r}_{i} + \bm{s})$. The components of this random vector are drawn from a uniform distribution in the interval $[-a_{0}/2, a_{0}/2]$. 
  
  \item Random velocities are drawn for each particle $\bm{v}_{i}^{\text{ran}}$ from the Maxwell-Boltzmann distribution with $\sqrt{k_{B}T_{0}/m_{0}}$ as standard deviation and zero mean.
  
  \item All particles are sorted in respective cells and cell-level quantities ($i.e.$ $N_{c}$, $\bm{r}_{c}$, $\bm{V}_{c}$, $\bm{V}_{c}^{\text{ran}}$, and $\bm{\Pi}_{c}$) are calculated.
  
  \item The velocity of fluid particles $\bm{v}_{i}$ is updated by performing the collision step following Eq.~\eqref{Collision}. The virtual particles are assigned the random velocity $\bm{v}_{i}^{\text{vp}} = \bm{v}_{i}^{\text{ran}}$.
  
  \item All the particles are shifted back to their original position as $\bm{r}_{i} \rightarrow (\bm{r}_{i} - \bm{s})$.
\end{enumerate}

\section{Mapping between MPCD units and physical system parameters} \label{sec:MPCD_mapping}
For many  nematic liquid crystals, the most important length scale is the nematic correlation length $\xi_{N} = \sqrt{L/(\alpha_{F} - 3 \beta_{F} S_{eq} + 18 \gamma_{F} S_{eq}^{2})}$ (with $S_{eq}$ as the scalar order parameter at equilibrium) which is the characteristic length scale over which the nematic order parameter varies significantly. The time scale over which the order parameter changes significantly is referred to as the nematic relaxation time $\tau_{N} = \mu_{1}\xi_{N}^{2}/L$. 
To capture the dynamics of the order parameter, we set $a_{0} = \xi_{N}$ and $t_{0} = \tau_{N}$. Next, we need to find $m_{0}$ and $k_{B}T_{0}$. As we want to reproduce the isotropic shear viscosity by  MPCD collisions, we use the definition of $\eta_{0}$ and the relation for isotropic viscosity and set $m_{0} = a_{0}t_{0}\beta_{4}/2 \eta_{\text{iso}}$. The thermal energy scale $k_{B}T_{0}$ can be obtained by simply using the definition of $t_{0}$ as $k_{B}T_{0} = m_{0}a_{0}^{2}/t_{0}^{2}$. Interested readers are referred to Padding and Louis \cite{padding2006hydrodynamic} for more extensive discussion on mapping between coarse-grained and physical systems. We take typical values of the phenomenological constants as $\alpha_{F} = - 3.333 \times 10^{-5} \ \text{J}/\text{m}^{3}$, $\beta_{F} = 8.888 \times 10^{-5} \ \text{J}/\text{m}^{3}$, and $\gamma_{F} = 4.444 \times 10^{-5} \ \text{J}/\text{m}^{3}$,  which yields $S_{eq} = 0.683$, $\xi_{N} = 1.099 \ \text{nm}$ and $\tau_{N} = 23.434 \ \text{ns}$. Now, using the previously mentioned scales, we can determine the parameters of Qian-Sheng theory for 5CB near 26$\celsius$ as $\mu_{1} = 107.991 \eta_{0}$, $\mu_{2} = -241.810 \eta_{0}$, $\beta_{1} = -16.699 \eta_{0}$, $\beta_{4} = 116.274 \eta_{0}$, $\beta_{5} = 182.498 \eta_{0}$, $\beta_{6} = -59.312 \eta_{0}$, $L = 107.991 k_{B}T_{0}/a_{0}$, $\alpha_{F} = -22.821 k_{B}T_{0}/a_{0}^{3}$, $\beta_{F} = 60.857 k_{B}T_{0}/a_{0}^{3}$ and $\gamma_{F} = 30.428 k_{B}T_{0}/a_{0}^{3}$. When the system is bounded by solid walls, we consider $W^{\text{vp}} = 10^{3} k_{B}T_{0}/a_{0}^{3}$ to impose strong anchoring condition.


%

\end{document}